\documentclass[a4paper,12pt]{article}
\usepackage{amsmath, amssymb, latexsym, amscd, amsthm,amsfonts,amstext}
\usepackage[mathscr]{eucal}
\usepackage{graphicx}
\usepackage{subfig}

\numberwithin{equation}{section}
\input{tcilatex}
\begin{document}

\title{Reconstruction procedures for two inverse scattering problems without
the phase information}
\author{Michael V. Klibanov$^{\ast }$ and Vladimir G. Romanov$^{\circ }$
\and $^{\ast }$Department of Mathematics and Statistics \and University of
North Carolina at Charlotte \and Charlotte, NC 28223, USA \and $^{\circ }$%
Sobolev Institute of Mathematics, Novosibirsk 630090, Russia \and E-mails:
mklibanv@uncc.edu and romanov@math.nsc.ru}
\maketitle

\begin{abstract}
This is a continuation of two recent publications of the authors \cite%
{KR,KRB} about reconstruction procedures for 3-d phaseless inverse
scattering problems. The main novelty of this paper is that, unlike \cite%
{KRB}, the Born approximation for the case of the wave-like equation is not
considered. It is shown here that the phaseless inverse scattering problem
for the 3-d wave-like equation in the frequency domain leads to the well
known Inverse Kinematic Problem. Uniqueness theorem follows. Still, since
the Inverse Kinematic Problem is very hard to solve, a linearization is
applied. More precisely, geodesic lines are replaced with straight lines. As
a result, an approximate explicit reconstruction formula is obtained via the
inverse Radon transform. The second reconstruction method is via solving a
problem of the integral geometry using integral equations of the Abel type.
\end{abstract}

% -------------------------------------------------------

\textbf{AMS classification codes:} 35R30.

\graphicspath{{Figures/}}

%
%
% \documentclass{amsart}
% %%%%%%%%%%%%%%%%%%%%%%%%%%%%%%%%%%%%%%%%%%%%%%%%%%%%%%%%%%%%%%
% \usepackage{amssymb}
% \usepackage[T1]{fontenc}
% \usepackage[latin1]{inputenc}
% \usepackage{graphicx}
% \usepackage{geometry}
% \usepackage{url}
% \usepackage{epsfig}
% % \usepackage{labelfig}
% \usepackage{verbatim}
% % \usepackage{umlaut}
% \usepackage{euscript}
% \usepackage{afterpage}
% \usepackage{graphics}
% \usepackage{amsmath}
% \usepackage{pst-plot}
% \usepackage{subfig}
%
% % \input{psfig.sty}
% \newtheorem{theorem}{Theorem}[section]
% \newtheorem{lemma}[theorem]{Lemma}
% \newtheorem{proposition}[theorem]{Proposition}
% \newtheorem{corollary}[theorem]{Corollary}
% \newtheorem{definition}[theorem]{Definition}
% \numberwithin{equation}{section}
% \newcommand{\norm}[1]{\left\Vert#1\right\Vert}
% \newcommand{\abs}[1]{\left\vert#1\right\vert}
% \newcommand{\set}[1]{\left\{#1\right\}}
% \newcommand{\Real}{\mathbb R}
% \newcommand{\eps}{\varepsilon}
% \newcommand{\To}{\longrightarrow}
% \newcommand{\bn}{\mathbf{n}}
% \def\OFEM{\Omega_{FEM}}
% \def\OFDM{\Omega_{FDM}}
% \newcommand\scal[1]{(\!(#1)\!)}
% \newcommand\bscal[1]{\big(\!\big(#1\big)\!\big)}
% % \input{tcilatex}
% \def\bR{\mathbb{R}}
% \def\bx{\mathbf{x}}
%
%
%
% \begin{document}
%
%
%
%
%

%
% \thispagestyle{plain}
%

\graphicspath{{FIGURES/}
{Figures/}
{FiguresJ/newfigures/}
{pics/}}

\section{Introduction}

\label{sec:1}

The Phaseless Inverse Scattering Problems (PISPs) arise in applications to
imaging of microstructures of sizes of the micron range of less (1 micron=1$%
\mu m=10^{-6}m).$ In particular, this includes imaging of nano structures of
sizes of hundreds of nanometers ($\approx 10^{-7}m)$ and biological cells
whose sizes are in the range of $\left( 5,100\right) \mu m$ \cite{PM,Bio}.
To image these objects, one needs to use either optical radiation with the
wavelength less than $1\mu m$ or the X-ray radiation. However, it is well
known that only the intensity (i.e. the square modulus) of the corresponding
complex valued wave field can be measured for such small wavelengths. The
phase cannot be measured, see, e.g. \cite{Dar,Die,Khach,Pet,Ruhl}.
Therefore, we arrive at the problem of the reconstruction of the scatterer
using only the intensity of the scattered wave field. In this case the
propagation of the wave field is governed by the wave-like PDE in the
frequency domain.

A similar problem, although for the Schr\"{o}dinger equation in the
frequency domain, arises in the quantum inverse scattering, where only the
differential scattering cross-section can be measured, which is actually the
square modulus of the solution of that equation, see, e.g. page 8 of \cite%
{Newton} and Chapter 10 of \cite{CS}. Unlike the wave-like equation (\ref%
{1.4}), in the case of the Schr\"{o}dinger equation the function $\beta
\left( x\right) $ (see (\ref{1.4})) is not multiplied by $k^{2},$ which
simplifies the problem. Note that, unlike PISPs, the conventional inverse
scattering theory is based on the assumption that both the intensity and the
phase of the complex valued wave field are measured, see, e.g. \cite%
{CS,Is,Newton,Nov1}.

A reconstruction procedure for a 3-d PISP for the wave-like PDE in the
frequency domain was proposed by the authors in \cite{KRB}. In \cite{KRB}
the linearization based on the Born approximation was used. However, the
Born approximation assumption is inconvenient, since it is actually assumed
there that $k^{2}\left\vert \beta \left( x\right) \right\vert <<1,$ where $%
\beta \left( x\right) $ is the unknown coefficient and $k$ is the frequency.
Hence, the Born approximation breaks down for $k>>1.$ On the other hand,
large values of $k$ are used in the reconstruction formula of \cite{KRB}.
Thus, the goal of this paper is to lift the assumption about the Born
approximation. We still linearize the problem.\ However, the \emph{main
novelty} here is that our linearization does not break down when $%
k\rightarrow \infty .$ We achieve this via an extensive use of the structure
of the fundamental solution of an associated hyperbolic equation with a
variable coefficient in its principal part.\ The latter is the main new
element here as compared with \cite{KR,KRB}. This structure was derived in 
\cite{R3}, also see \cite{R2}.

The first uniqueness result for the PISP was proven in \cite{KS} for the 1-d
case with some follow up publications \cite{AS,NHS}.\ Uniqueness for the 3-d
case was proven in \cite{KSIAP,AML,AA}. However, proofs in these references
are not constructive. In fact, besides just uniqueness only it is desirable
to develop rigorous numerical methods for the phase reconstruction. Many
heuristic attempts were made by physicists to reconstruct the phase, see,
e.g. \cite{Dar,Die,Khach,Pet,Ruhl}.

However, rigorous numerical methods for PISPs were derived only very
recently by the authors and Novikov \cite{KR,KRB,Nov3,Nov4}. In \cite{KR} a
3-d PISP for the Schr\"{o}dinger equation was considered and an explicit
reconstruction formula was derived, which is based on the inverse Radon
transform, see, e.g. \cite{Nat} for this transform. Thus, a long standing
problem posed in 1977 in Chapter 10 of \cite{CS} was addressed in \cite{KR}
for the first time. In \cite{KRB} the result of \cite{KR} was extended to
the case of the wave-like equation (\ref{1.4}) using the Born approximation
assumption.

While only unknown scatterers are involved in measurements in \cite{KR,KRB},
the reconstruction formula of \cite{Nov3} requires the involvement of two
more known scatterers. A quite general reconstruction procedure of \cite%
{Nov4} is using measurements of the intensity of the full wave field on at
least two spheres in the far field zone.\ The latter is unlike \cite%
{KR,KRB,Nov3}, where the intensity of only the scattered wave field is
measured on just one surface and the far field approximation is not used.
Note that usually the intensity of the scattered rather than the full wave
field is measured. This is because measurements are conducted only outside
of the brightening area since detectors are \textquotedblleft burned" in the
brightening area.\ For example, in images on page 22 of \cite{Die} the areas
with brightening are depicted in the red color. On the other hand, outside
of the brightening area the intensity of the full wave field is well
approximated by the intensity of the scattered wave field. While works \cite%
{KSIAP,AML,AA,KR,KRB,Nov3,Nov4} are concerned with the reconstructions of
coefficients of PDEs from the phaseless scattered data, publications \cite%
{Iv1,Iv2} consider numerical methods of the reconstruction of shapes of
obstacles from the phaseless scattered data.

We present here two reconstruction methods for two PISPs. In both cases we
use measurements of the intensity on a single sphere on many frequencies and
the far field approximation is not used. First, we reduce our PISP to the
Inverse Kinematic Problem \cite{LRV,R1,R2}. This leads to a new uniqueness
result, which is based on two items: (1) a uniqueness theorem of \cite{R2}
for the Inverse Kinematic Problem and (2) our reconstruction procedure.
Next, we linearize the Inverse Kinematic Problem, as in \cite{LRV,R1,R2},
and reconstruct the unknown coefficient via the inverse Radon transform, as
in \cite{KR,KRB}. In the second approach we obtain after that linearization
a problem of the integral geometry and solve it explicitly via solving
certain integral equations of the Abel type. In our linearization we assume,
similarly with \cite{LRV,R1,R2}, that certain integrals over geodesic lines
are actually integrals over straight lines.

As it is often happen to other reconstruction procedures (see, e.g. \cite%
{Nov3,Nov4}), our two reconstruction procedures cannot be considered as
algorithms yet. In other words, they cannot be considered as sequences of
steps leading to the numerical solutions. To turn them in algorithms, our
reconstruction steps must be complemented by some regularization steps.
Although the latter is possible, we leave this to future numerical
publications for brevity.

In section 2 statements of problems under consideration are presented. In
section 3 we consider an auxiliary Cauchy problem for a hyperbolic equation,
which is important for our study. In section 4 we present a reconstruction
method via the inversion of the Radon transform. Finally in section 5 we
present the second reconstruction method via solution of a problem of the
integral geometry.

\section{Problem statement}

\label{sec:2}

Let $B>0$ be a number. Let $\Omega =\{|x|<R\}\subset \mathbb{R}^{3}$ be the
ball of the radius $R<B$ with the center at $\{0\}$ Denote $Y=\left\{
|x|<B\right\} ,S=\left\{ \left\vert x\right\vert =B\right\} .$ Let $%
n(x),x\in \mathbb{R}^{3}$ be a real valued function satisfying the following
conditions 
\begin{equation}
n\in C^{15}(\mathbb{R}^{3}),\quad n^{2}(x)=1+\beta (x),  \label{1.1}
\end{equation}%
\begin{equation}
\beta (x)\geq 0,\text{ }\beta (x)=0\quad \text{for }\>x\in \mathbb{R}%
^{3}\setminus \Omega .  \label{1.3}
\end{equation}%
The smoothness requirement imposed on the function $n(x)$ is clarified in
the proof of Theorem 1 in subsection 3.1. The conformal Riemannian metric
generated by $n(x)$ is 
\begin{equation}
d\tau =n(x)\left\vert dx\right\vert ,|dx|=\sqrt{%
(dx_{1})^{2}+(dx_{2})^{2}+(dx_{3})^{2}}.  \label{1.31}
\end{equation}%
\ Below we impose the following Assumption:

\textbf{Assumption}. We assume that geodesic lines of the metric (\ref{1.31}%
) satisfy the regularity condition, i.e. for each two points $x,y\in \mathbb{%
R}^{3}$ there exists a single geodesic line $\Gamma \left( x,y\right) $
connecting these points.

It is well known from the Hadamard-Cartan theorem \cite{Ball} that in any
simply connected complete manifold with a non positive curvature each two
points can be connected by a single geodesic line. The manifold $(\Omega ,n)$
is called the manifold of a non positive curvature, if the section
curvatures $K(x,\sigma )\leq 0$ for all $x\in \Omega $ and for all
two-dimensional planes $\sigma $. A sufficient condition for $K(x,\sigma
)\leq 0$ was derived in \cite{R4} 
\begin{equation*}
\sum_{i,j=1}^{3}\frac{\partial ^{2}\ln n(x)}{\partial x_{i}\partial x_{j}}%
\xi _{i}\xi _{j}\geq 0,\>\forall x,\xi \in \mathbb{R}^{3}.
\end{equation*}

For $x,y\in \mathbb{R}^{3},$ let $\tau (x,y)$ be the solution to the problem 
\begin{equation}
|\nabla _{x}\tau (x,y)|^{2}=n^{2}(x),\quad \tau (x,y)=O\left( \left\vert
x-y\right\vert \right) ,\text{ }\>\mathrm{as}\text{ }\>y\rightarrow x.
\label{1.9}
\end{equation}%
Let $d\sigma $ be the euclidean arc length of the geodesic line $\Gamma
\left( x,y\right) .$ Then 
\begin{equation}
\tau \left( x,y\right) =\dint\limits_{\Gamma \left( x,y\right) }n\left( \xi
\right) d\sigma .  \label{1.90}
\end{equation}%
Hence, $\tau (x,y)$ is the travel time between points $x$ and $y$ due to the
Riemannian metric (\ref{1.31}). Due the Assumption, $\tau (x,y)$ is a
single-valued function of both points $x$ and $y$ in $\mathbb{R}^{3}\times 
\mathbb{R}^{3}$.

We consider the following equation 
\begin{equation}
\Delta u+k^{2}n^{2}(x)u=-\delta (x-y),\quad x\in \mathbb{R}^{3},  \label{1.4}
\end{equation}%
where the Laplace operator is taken with respect to $x$, the frequency $k>0$
is a positive real number and $y\in \mathbb{R}^{3}$ is the source position.
Naturally, we assume that the function $u(x,k,y)$ satisfies the radiation
condition 
\begin{equation}
\frac{\partial u}{\partial r}+iku=o(r^{-1})\>\text{ }\mathrm{as}\text{ }%
\>r=|x-y|\rightarrow \infty .  \label{1.5}
\end{equation}%
Denote $u_{0}(x,k,y)$ the solution of the problem (\ref{1.4}), (\ref{1.5})
for the case $n(x)\equiv 1.$ Then $u_{0}$ is the incident spherical wave,%
\begin{equation*}
u_{0}(x,k,y)=\frac{\exp \left( -ik\left\vert x-y\right\vert \right) }{4\pi
\left\vert x-y\right\vert }.
\end{equation*}%
Let $u_{sc}(x,k,y)$ be the scattered wave, which is due to the presence of
scatterers, in which $n(x)\neq 1$. Then 
\begin{equation}
u_{sc}(x,k,y)=u(x,k,y)-u_{0}(x,k,y)=u(x,k,y)-\frac{\exp \left( -ik\left\vert
x-y\right\vert \right) }{4\pi \left\vert x-y\right\vert }.  \label{1.6}
\end{equation}%
Combining Theorem 8.7 of \cite{CK} with Theorem 6.17 of \cite{GT} and taking
into account that $n\in C^{15}(\mathbb{R}^{3}),$ we obtain that the problem (%
\ref{1.4}), (\ref{1.5}) has unique solution $u\in C^{16+\alpha }\left(
\left\vert x-x_{0}\right\vert \geq \eta \right) ,\forall \eta >0$ for any $%
\alpha \in \left( 0,1\right) .$ Here $C^{k+\alpha }$ are H\"{o}lder spaces.

We consider the following two Phaseless Inverse Scattering Problems (PISPs):

\textbf{PISP1}. \emph{Suppose that the following function }$f_{1}\left(
x,k,y\right) $\emph{\ is known}%
\begin{equation}
f_{1}\left( x,k,y\right) =\left\vert u_{sc}(x,k,y)\right\vert ^{2},\>\forall
\left( x,y\right) \in S\times S,\>\forall k\geq k_{0},  \label{1.51}
\end{equation}%
\emph{where }$k_{0}=const.>0.$\emph{\ Determine the function }$\beta \left(
x\right) $\emph{\ satisfying conditions (\ref{1.1}), (\ref{1.3}).}

\textbf{PISP2}. \emph{Suppose that the following function }$f_{2}\left(
x,k,y\right) $\emph{\ is known}%
\begin{equation}
f_{2}\left( x,k,y\right) =\left\vert u(x,k,y)\right\vert ^{2},\>\forall
\left( x,y\right) \in S\times S,\>\forall k\geq k_{0},  \label{1.52}
\end{equation}%
\emph{where }$k_{0}=const.>0.$\emph{\ Determine the function }$\beta \left(
x\right) $\emph{\ satisfying conditions (\ref{1.1}), (\ref{1.3}).}

In both these problems we do not take into account the fact that the
intensity of the wave field can be measured only outside of the brightening
area (see Introduction). This case is a more difficult one than we now
consider and we hope to study it in the future.

Below we present a reconstruction method for the PISP1 in a linearized
case.\ This method is based on the inverse Radon transform. In addition, we
present a different reconstruction procedure both for the PISP1 and for the
PISP2 (both are linearized) via solving a problem of the integral geometry.
Note that while the inverse Radon transform is applicable to the PISP1, it
is inapplicable to the PISP2. On the other hand, the reconstruction methods
based on the integral geometry work for both these problems.

\textbf{Remark}. In fact, it follows from our reconstruction procedures
that, in the linearized case, it is sufficient to know functions $%
f_{1}\left( x,k,y\right) ,f_{2}\left( x,k,y\right) $ only for all $\left(
x,y\right) \in \left( S\cap \left\{ x_{3}=z\right\} \right) \times \left(
S\cap \left\{ x_{3}=z\right\} \right) $ for each $z\in \left( -R,R\right) $
and for all $k\geq k_{0}.$ This case of data collection is more economical
than the one of (\ref{1.51}), (\ref{1.52}). In this case the function $\beta
\left( x\right) $ is reconstructed separately in each 2-d cross-section $%
\Omega \cap \left\{ x_{3}=z\right\} $ of the domain $\Omega .$

\section{The auxiliary Cauchy problem for a hyperbolic equation}

\label{sec:3}

Consider the following Cauchy problem 
\begin{equation}
n^{2}(x)v_{tt}=\Delta v+\delta (x-y,t),\quad x\in \mathbb{R}^{3},t>0,
\label{3.1}
\end{equation}
\begin{equation}
v\left( x,0\right) =v_{t}\left( x,0\right) =0.  \label{3.2}
\end{equation}

\subsection{The form of the solution of the problem (\protect\ref{3.1}), (%
\protect\ref{3.2})}

\label{sec:3.1}

Let $\zeta =(\zeta _{1},\zeta _{2},\zeta _{3})$, $\zeta =\zeta (x,y)$ be
geodesic coordinates of a variable point $x$ with respect to a fixed point $%
y $ in the above Riemannian metric (\ref{1.31}). By the above Assumption,
there exists a one-to-one correspondence $x\Leftrightarrow \zeta $ for any
fixed $y$. Therefore, for any fixed point $y$ the function $\zeta =\zeta
(x,y)$ has the inverse function $x=f\left( \zeta ,y\right) $ which
determines the geodesic line $\Gamma (x,y)=\{\xi :\xi =f\left( s\zeta
_{0},y\right) ,s\in \lbrack 0,\tau (x,y)]\}$, where $\zeta _{0}$ is the
vector which is tangent to $\Gamma (x,y)$ at the point $y$, is directed
towards the point $x$ and also $|\zeta _{0}|=n^{-1}(y)$. Moreover, by the
formula (2.2.28) in the book \cite{R3} the function $\zeta (x,y)$ can be
expressed via the function $\tau (x,y)$ as 
\begin{equation}
\zeta (x,y)=-\frac{1}{2n^{2}(y)}\nabla _{y}\tau ^{2}(x,y).  \label{1.9a}
\end{equation}%
Note that in our case one should take in the formula (2.2.28) of \cite{R3} $%
A(y)=n^{-2}(y)I$, where $I$ is the unit matrix. Also, since by (\ref{1.1}) $%
n(x)\in C^{15}(\mathbb{R}^{3})$ and the Assumption holds, then $\tau
^{2}(x,y)\in C^{15}(\mathbb{R}^{3}\times \mathbb{R}^{3})$, $k\geq 2$, and $%
\zeta (x,y)\in C^{14}(\mathbb{R}^{3}\times \mathbb{R}^{3})$ \cite{R3}.
Consider the Jacobian $J(x,y)$, 
\begin{equation}
J(x,y)=\mathrm{\det }\frac{\partial \zeta }{\partial x}.  \label{1.91}
\end{equation}%
By the formula (2.2.18) in \cite{R3} $\zeta =x-y+O(|x-y|^{2})$ as $%
x\rightarrow y$. Hence, 
\begin{equation*}
\left. \frac{\partial \zeta _{i}}{\partial x_{j}}\right\vert
_{x=y}=\lim_{x_{j}-y_{j}=h\rightarrow 0}\frac{h\delta _{ij}+O(h^{2})}{h}%
=\delta _{ij},\>i,j=1,2,3,
\end{equation*}%
where $\delta _{ij}$ is the Kronecker delta. Hence, $J(y,y)=1$. Next, the
smoothness of the function $\zeta (x,y)$ with respect to $x$ implies that
the function $J(x,y)$ is continuous with respect to $x$. Further, since the
correspondence $x\Leftrightarrow \zeta $ is one-to-one and differentiable,
then (\ref{1.91}) implies that\textbf{\ }$J(x,y)\neq 0,$ $\forall \left(
x,y\right) \in \mathbb{R}^{3}\times \mathbb{R}^{3}.$ Hence, 
\begin{equation}
J(x,y)>0,\forall \left( x,y\right) \in \mathbb{R}^{3}\times \mathbb{R}^{3}.
\label{1.92}
\end{equation}

For any $y\in \mathbb{R}^{3}$ and for any $T>0$ consider two domains $D(y,T)$
and $D^{\ast }(y,T),$%
\begin{equation*}
D(y,T)=\{(x,t):0<\>t\leq T-|x-y|\}\text{ and }D^{\ast }(y,T)=\{(x,t):\tau
(x,y)\leq t\leq T-\tau (x,y)\}.
\end{equation*}

\textbf{Theorem 1.} \emph{Let} \emph{the function }$n(x)$\emph{\ satisfies
conditions (\ref{1.1}) and (\ref{1.3})} \emph{\ and let the Assumption
holds. Then for any }$T>0$ \emph{\ and for any fixed }$y\in \mathbb{R}^{3}$ 
\emph{\ there exists unique solution } $v(x,t,y)$ \emph{\ to the problem (%
\ref{3.1}), (\ref{3.2}), which can be represented in the domain } $D(y,T)$ 
\emph{\ in the form }%
\begin{equation}
v(x,t,y)=A(x,y)\delta (t-\tau (x,y))+\hat{v}(x,t,y)H(t-\tau (x,y)),
\label{1.10}
\end{equation}%
\emph{\ where } $A(x,y)>0$ \emph{\ is given by the formula }%
\begin{equation}
A(x,y)=\frac{n^{2}(y)\sqrt{J(x,y)}}{4\pi n(x)\tau (x,y)},  \label{1.11}
\end{equation}%
$H(t)$ \emph{is the Heaviside function and the function }$\hat{v}(x,t,y)\in
C^{2}\left( D^{\ast }(y,T)\right) $\emph{.}

\textbf{Proof}. We note first that we can use $\sqrt{J(x,y)}$ in (\ref{1.11}%
) since by (\ref{1.92}) $J(x,y)>0$. Rewrite the problem (\ref{3.1}), (\ref%
{3.2}) in the form 
\begin{equation}
v_{tt}-\mathrm{div}\left( n^{-2}(x)\nabla v\right) +\nabla n^{-2}(x)\cdot
\nabla v=n^{-2}(y)\delta (x-y,t),\quad v|_{t<0}\equiv 0,  \label{4.1}
\end{equation}%
We now use results obtained in the book \cite{R3}. Represent the solution to
the problem (\ref{4.1}) for $t\geq 0$ in the form 
\begin{equation*}
v(x,t,y)=
\end{equation*}%
\begin{equation}
\frac{1}{n^{2}(y)}\Bigg[a_{-1}(x,y)\delta (t^{2}-\tau
^{2}(x,y))+\sum\limits_{k=0}^{s}a_{k}(x,y)\theta _{k}(t^{2}-\tau
^{2}(x,y))+w_{s}(x,t,y)\Bigg],  \label{4.2}
\end{equation}%
where the integer $s>1$ is chosen below and functions $\theta _{k}(t)$ are
defined as 
\begin{equation*}
\theta _{k}(t)=\frac{t^{k}}{k!}H(t),\quad k=0,1,2,\ldots ,s.
\end{equation*}%
Coefficients $a_{k}(x,y)$ in (\ref{4.2}) are defined by the following
formulas 
\begin{eqnarray}
&&a_{-1}(x,y)=\frac{\sqrt{J(x,y)}n^{4}(y)}{2\pi n(x)},  \label{4.4} \\
&&a_{r}(x,y)=\frac{a_{-1}(x,y)}{4\tau ^{r-1}(x,y)}\int\limits_{\Gamma
(x,y)}\tau ^{r}(\xi ,y)(a_{-1}(\xi ,y))^{-1}n^{-1}(\xi )\Delta _{\xi
}a_{r-1}(\xi ,y)d\sigma ,\>r\geq 0,\qquad  \label{4.40}
\end{eqnarray}%
where $\xi $ is a variable point on this geodesic line $\Gamma (x,y)$. We
have derived formulae (\ref{4.4}), (\ref{4.40}) from formulas
(2.2.41)-(2.2.44) of the book \cite{R3}, in which the notation $\alpha _{r}$
was used for $a_{r}$ and $d\tau ^{\prime }=n(\xi )d\sigma $. We took $m=1$
in the latter formulae.

The residual function $w_{s}(x,t,y)$ in (\ref{4.2}) is the solution of the
following problem 
\begin{equation}
\partial ^{2}w_{s}-n^{-2}(x)\Delta w_{s}=F_{s}(x,t,y),\quad w|_{t<0}\equiv 0,
\label{4.2a}
\end{equation}%
\begin{equation}
F_{s}(x,t,y)=n^{-2}(x)(\Delta a_{s}(x,y))\theta _{s}(t^{2}-\tau ^{2}(x,y)).
\label{4.2b}
\end{equation}

Let $k>1$ be a sufficiently large integer which is chosen below. If the
function $n(x)\in C^{k}(\mathbb{R}^{3}),$ then $\tau ^{2}(x,y)\in C^{k}(%
\mathbb{R}^{3}\times \mathbb{R}^{3})$ (see \cite{R3}, p. 27). Therefore it
follows from (\ref{4.4}) and (\ref{4.40}) that 
\begin{equation*}
a_{-1}(x,y)\in C^{k-2}(\mathbb{R}^{3}\times \mathbb{R}^{3}),\quad
a_{r}(x,y)\in C^{k-4-2r}(\mathbb{R}^{3}\times \mathbb{R}^{3}),\>r=0,1,\ldots
,s.
\end{equation*}%
Hence, the function $F_{s}(x,t,y)\in C^{l}(D(y,T))$, where $l=\min
(m-6-2s,s) $ and this function vanishes for $t\leq \tau (x,y)$. Hence, $%
w_{s}(x,t,y)\equiv 0$ for $t\leq \tau (x,y)$.

Using theorems 3.2, 4.1, Corollary 4.2 and energy estimates of Chapter 4 of 
\cite{Lad}, one can easily prove that $w_{s}\in C^{l-1}\left( D(y,T)\right)
, $ see, e.g. Theorem 2.2 of \cite{KTat} for a similar result. Choosing $%
m=15 $ and $s=3$, we obtain $l=3$. Thus, $w_{3}\in C^{2}(D(y,T))$. Since by (%
\ref{1.11}) and (\ref{4.4}) $A(x,y)=a_{-1}(x,y)/(2n^{2}(y)\tau (x,y))$, then 
$A(x,y)>0$ and $A(x,y)\in C^{13}(\mathbb{R}^{3}\times \mathbb{R}^{3})$.
Moreover, setting 
\begin{equation*}
\hat{v}(x,t,y)=\frac{1}{n^{2}(y)}\Bigg[\sum\limits_{r=0}^{3}a_{r}(x,y)\frac{%
(t^{2}-\tau ^{2}(x,y))^{r}}{r!}+w_{3}(x,t,y)\Bigg],\>(x,y)\in D^{\ast }(T,y),
\end{equation*}%
we see that formula (\ref{4.2}) coincides with formula (\ref{1.10}) and also
that the function $\hat{v}(x,t,y)\in C^{2}(D^{\ast }(T,y))$. $\square $

\subsection{Connection with the problem (\protect\ref{1.4}), (\protect\ref%
{1.5})}

\label{sec:3.2}

Let $\Phi \subset \mathbb{R}^{3}$ be an arbitrary bounded domain. Lemma 6 of
Chapter 10 of the book \cite{V} as well as Remark 3 after that lemma
guarantee that functions $\partial _{t}^{k}v(x,t,\nu ),k=0,1,2$ and $\Delta
_{x}v(x,t,\nu )$ decay exponentially as $t\rightarrow \infty ,$ as longs as $%
x$ reminds in the domain $K$. In other words, there exist constants $%
M=M\left( \Phi ,\beta \right) >0,c=c\left( \Phi ,\beta \right) >0$ such that 
\begin{equation}
\left\vert \partial _{t}^{k}v\left( x,t,y\right) \right\vert ,\left\vert
\Delta v(x,t,y)\right\vert \leq Me^{-c\,t}\text{ for all }\>t\geq t_{0}\>%
\text{ and for all }\>x\in \Phi ,  \label{5.1}
\end{equation}%
where $t_{0}=t_{0}\left( \Phi ,\beta \right) =const.>0.$ Hence, one can
consider the Fourier transform $V\left( x,k,y\right) $ of the function $v,$%
\begin{equation}
V(x,k,y)=\dint\limits_{0}^{\infty }v\left( x,t,y\right) \exp \left(
-ikt\right) dt.  \label{5.2}
\end{equation}%
Next, Theorem 3.3 of \cite{V1} and Theorem 6 of Chapter 9 of \cite{V}
guarantee that $V(x,k,y)=u(x,k,y),$ where the function $u(x,k,y)$ is the
above solution of the problem (\ref{1.4}), (\ref{1.5}).

Comparing (\ref{5.1}) and (\ref{5.2}) with (\ref{1.10}), integrating by
parts in (\ref{5.2}) and also using (\ref{1.6}), we obtain%
\begin{equation}
u(x,k,y)=\exp (-ik\tau (x,y))\left[ A(x,y)+O\left( \frac{1}{k}\right) \right]
,k\rightarrow \infty ,  \label{5.3}
\end{equation}%
\begin{equation}
u_{sc}(x,k,y)=A(x,y)\exp (-ik\tau (x,y))-\frac{\exp \left( -ik\left\vert
x-y\right\vert \right) }{4\pi \left\vert x-y\right\vert }+O\left( \frac{1}{k}%
\right) ,k\rightarrow \infty .  \label{5.4}
\end{equation}%
Hence, (\ref{1.51}), (\ref{1.52}), (\ref{5.3}) and (\ref{5.4}) imply that
for $\left( x,y\right) \in S\times S$%
\begin{equation*}
f_{1}\left( x,k,y\right) =
\end{equation*}%
\begin{equation}
A^{2}(x,y)+\frac{1}{16\pi ^{2}\left\vert x-y\right\vert ^{2}}-\frac{A(x,y)}{%
2\pi \left\vert x-y\right\vert }\cos \left[ k\left( \tau (x,y)-\left\vert
x-y\right\vert \right) \right] +O\left( \frac{1}{k}\right) ,k\rightarrow
\infty ,  \label{5.5}
\end{equation}%
\begin{equation}
f_{2}\left( x,k,y\right) =A^{2}(x,y)+O\left( \frac{1}{k}\right)
,k\rightarrow \infty .  \label{5.6}
\end{equation}

\section{Approximate solution of the linearized PISP1 via the inverse Radon
transform}

\label{sec:4}

\subsection{Reconstruction of the function $\protect\tau (x,y)$ for $\left(
x,y\right) \in S\times S$}

\label{sec:4.1}

Below in this subsection $k\geq k_{1}>k_{0},$ where $k_{1}>>1$ is a
sufficiently large number. We now fix the point $\left( x,y\right) \in
S\times S$ and consider $f_{1}(x,k,y)$ as the function of $k$ for $k\geq
k_{1}$. It is possible to figure out whether or not $\tau \left( x,y\right)
=\left\vert x-y\right\vert .$ Indeed, it follows from (\ref{5.6}) that $\tau
(x,y)=|x-y|$ if and only if $\lim_{k\rightarrow \infty }f_{1}(x,k,y)$
exists. Ignoring in (\ref{5.5}) the term $O\left( k^{-1}\right) ,$ we obtain
for $k\geq k_{1}$ 
\begin{equation}
f_{1}\left( x,k,y\right) =A^{2}(x,y)+\frac{1}{16\pi ^{2}\left\vert
x-y\right\vert ^{2}}-\frac{A(x,y)}{2\pi \left\vert x-y\right\vert }\cos %
\left[ k\left( \tau (x,y)-\left\vert x-y\right\vert \right) \right] .
\label{6.1}
\end{equation}%
In (\ref{6.1}) we use \textquotedblleft $=$" instead of \textquotedblleft $%
\approx $". By (\ref{6.1}) there exists a number $k_{2}\geq k_{1}$ such that 
\begin{equation}
f_{1}^{\ast }(x,y)=f_{1}\left( x,k_{2},y\right) =\max_{k\geq
k_{1}}f_{1}\left( x,k,y\right) =\left( A(x,y)+\frac{1}{4\pi |x-y|}\right)
^{2}  \label{6.2}
\end{equation}%
Hence, we find the number $A(x,y)$ as 
\begin{equation}
A(x,y)=\sqrt{f_{1}^{\ast }(x,y)}-\frac{1}{4\pi |x-y|}.  \label{6.20}
\end{equation}

Assume that $\tau (x,y)\neq |x-y|.$ Then the positivity of the function $%
\beta \left( x\right) $ and (\ref{1.90}) imply that $\tau (x,y)>\left\vert
x-y\right\vert .$ Choose the number $k_{3}>k_{2}$ such that 
\begin{equation}
k_{3}=\min \left\{ k:k>k_{2},f_{1}\left( x,k,y\right) =f_{1}^{\ast
}(x,y)\right\} .  \label{6.3}
\end{equation}%
Then (\ref{6.1})-(\ref{6.3}) imply that 
\begin{equation*}
k_{3}\left( \tau (x,y)-\left\vert x-y\right\vert \right) =k_{2}\left( \tau
(x,y)-\left\vert x-y\right\vert \right) +2\pi .
\end{equation*}%
Therefore, we reconstruct the number $\tau (x,y)$ as 
\begin{equation}
\tau (x,y)=|x-y|+\frac{2\pi }{k_{3}-k_{2}}.  \label{6.4}
\end{equation}

Next, we should reconstruct the function $\beta \left( x\right) ,$ which is
done below. To do this, we consider a linearization.\ However, we now can
formulate uniqueness theorem for PISP1 without the linearization assumption.
This is because the knowledge of the function $\tau (x,y)$ for all $\left(
x,y\right) \in S\times S,$ which follows from (\ref{6.4}), is equivalent to
the so-called Inverse Kinematic Problem.\ This problem was studied in \cite%
{LRV,R1,R2}. In particular, Theorem 3.4 of Chapter 3 of \cite{R2} claims
uniqueness of the reconstruction of the function $n\left( x\right) $ from
the knowledge of the function $\tau (x,y)$ for all $\left( x,y\right) \in
S\times S.$ Thus, our Theorem 2 follows immediately from Theorem 3.4 of
Chapter 3 of \cite{R2} as well as from (\ref{6.4}).

\textbf{Theorem 2} (uniqueness). \emph{Suppose that two functions }$%
n_{1}\left( x\right) ,n_{2}\left( x\right) $\emph{\ satisfying conditions (%
\ref{1.1}), (\ref{1.3}) also satisfy Assumption of section 2. In addition,
assume that these two functions generate the same function }$f_{1}\left(
x,k,y\right) $\emph{\ in (\ref{1.51}) and that the function }$f_{1}\left(
x,k,y\right) $\emph{\ has the form (\ref{6.1}), i.e. the term }$O\left(
1/k\right) $\emph{\ in (\ref{5.5}) is dropped. Then }$n_{1}\left( x\right)
\equiv n_{2}\left( x\right) $\emph{.}

\subsection{Reconstruction of the function $\protect\beta \left( x\right) $}

\label{sec:4.2}

Assume that 
\begin{equation}
||\beta ||_{C^{2}\left( \overline{\Omega }\right) }<<1.  \label{6.6}
\end{equation}%
Then the linearization of the function $\tau \left( x,y\right) $ with
respect to the function $\beta $ leads to 
\begin{equation}
\tau \left( x,y\right) =|x-y|+\dint\limits_{L\left( x,y\right) }\beta \left(
\xi \right) d\sigma ,  \label{6.7}
\end{equation}%
where $L\left( x,y\right) $ is the segment of the straight line connecting
points $x$ and $y$ and $d\sigma $ is its arc length, see Theorem 11 in
Chapter 3 of \cite{LRV} as well as \S 5 in Chapter 2 of \cite{R1} and \S 4
in Chapter 3 of \cite{R2}. To be precise, in (\ref{6.7}) one should have
\textquotedblleft $\approx $" instead of \textquotedblleft $=$". Since the
function $\tau \left( x,y\right) $ was approximately reconstructed via (\ref%
{6.4}), then we obtain from (\ref{6.7}) that the following function $h\left(
x,y\right) =\tau \left( x,y\right) -|x-y|$ is known%
\begin{equation}
h\left( x,y\right) =\dint\limits_{L\left( x,y\right) }\beta \left( \xi
\right) ds,\>\forall \left( x,y\right) \in S\times S.  \label{6.8}
\end{equation}

For any number $z\in \mathbb{R}$ consider the plane $P_{z}=\left\{
x_{3}=z\right\} .$ Consider the disk $Q_{z}=\overline{Y}\cap P_{z}$ and let $%
S_{z}=S\cap P_{z}$ be its boundary. Clearly $Q_{z}\neq \varnothing $ for $%
z\in \left( -B,B\right) $ and $Q_{z}=\varnothing $ for $\left\vert
z\right\vert \geq B.$ Denote $0_{z}=\left( 0,0,z\right) \in Q_{z}$ the
orthogonal projection of the origin on the plane $P_{a}.$ For an arbitrary $%
z\in \left( -B,B\right) $ denote $B_{z}=\sqrt{B^{2}-z^{2}}$ the radius of
the circle $S_{z}.$ We have 
\begin{equation*}
Y=\dbigcup\limits_{z=-B}^{B}Q_{z},\partial
Y:=S=\dbigcup\limits_{z=-B}^{B}S_{z},\Omega \subset Y.
\end{equation*}

We now introduce some notations of the Radon transform, which we take from
the book \cite{Nat}. Since our reconstruction formula is based on the
inversion of the two-dimensional\ Radon transform, we now parametrize $%
L\left( x,y\right) $ in the conventional way of the parametrization of the
Radon transform \cite{Nat}. Let $\nu $ be the unit normal vector to $L\left(
x,y\right) $ lying in the plane $P_{z}$ and pointing outside of the point $%
0_{z}.$ Let $\alpha \in \left( 0,2\pi \right] $ be the angle between $\nu $
and the $x_{1}-$axis. Then $\nu =\nu \left( \alpha \right) =\left( \cos
\alpha ,\sin \alpha \right) $ (it is convenient here to discount the third
coordinate of $\nu ,$ which is zero). Let $s$ be the signed distance between 
$L\left( x,y\right) $ and the point $0_{z}$ (page 9 of \cite{Nat}). It is
clear that there exists a one-to-one correspondence between pairs $\left(
x,y\right) $ and $\left( \nu \left( \alpha \right) ,s\right) ,$%
\begin{equation}
\left( x,y\right) \Leftrightarrow \left( \nu \left( \alpha \right) ,s\right)
,\left( x,y\right) \in S_{z}\times S_{z},\alpha =\alpha \left( x,y\right)
\in \left( 0,2\pi \right] ,s=s\left( x,y\right) \in \left(
-B_{z},B_{z}\right) .  \label{6.9}
\end{equation}%
Hence, we can write 
\begin{equation}
L\left( x,y\right) =\left\{ r_{a}=\left( r_{1},r_{2},a\right) :\left\langle
r,\nu \left( \alpha \right) \right\rangle =s\right\} ,  \label{6.10}
\end{equation}%
where $r=\left( r_{1},r_{2}\right) \in \mathbb{R}^{2},\left\langle
,\right\rangle $ is the scalar product in $\mathbb{R}^{2}$ and parameters $%
\alpha =\alpha \left( x,y\right) $ and $s=s\left( x,y\right) $ are defined
as in (\ref{6.9}).

Consider an arbitrary function $g=g\left( r\right) \in C^{4}\left(
P_{z}\right) $ such that $g\left( r\right) =0$ for $y\in $ $P_{z}\diagdown
Q_{z}.$ Hence, 
\begin{equation}
\dint\limits_{L\left( x,y\right) }g\left( r\right) d\sigma
=\dint\limits_{\left\langle r,\nu \left( \alpha \right) \right\rangle
=s}g\left( r\right) d\sigma ,\forall \left( x,y\right) \in S_{z}\times S_{z},
\label{6.11}
\end{equation}%
where $\alpha =\alpha \left( x,y\right) ,s=s\left( x,y\right) $ are as in (%
\ref{6.9}). The parametrization of $L\left( x,y\right) $ in (\ref{6.11}) is
as in (\ref{6.10}). Therefore, using (\ref{6.9})-(\ref{6.11}), we can define
the 2-d Radon transform $Rg$ of the function $g$ as 
\begin{equation*}
\left( Rg\right) \left( x,y\right) =\left( Rg\right) \left( \alpha ,s\right)
=\dint\limits_{\left\langle r,\nu \left( \alpha \right) \right\rangle
=s}g\left( r\right) d\sigma ,\>\forall \left( x,y\right) \in S_{z}\times
S_{z}.
\end{equation*}

Hence, by (\ref{6.8}) and (\ref{6.9}) 
\begin{equation}
h\left( x,y\right) =\left( R\beta \right) \left( \alpha ,s\right)
=\dint\limits_{\left\langle r,\nu \left( \alpha \right) \right\rangle
=s}\beta \left( r,a\right) d\sigma ,\>\forall \left( x,y\right) \in
S_{z}\times S_{z},\forall z\in \left( -B,B\right) .  \label{6.12}
\end{equation}%
Finally, since the function $h\left( x,y\right) $ is known, then (\ref{6.12}%
) leads to the following reconstruction formula%
\begin{equation}
\beta \left( r_{1},r_{2},a\right) =R^{-1}\left( h\left( x,y\right) \right)
\left( r_{1},r_{2},a\right) ,\left( x,y\right) \in S_{z}\times S_{z},\forall
z\in \left( -B,B\right) .  \label{6.120}
\end{equation}%
Formula (\ref{6.120}) is our \emph{final reconstruction result} for PISP1
via the inverse Radon transform. Since the formula for $R^{-1}$ is well
known \cite{Nat}, we do not specify it here for brevity.

We now formulate uniqueness result for the linearized PISP1. This result
follows immediately from (\ref{6.12}).

\textbf{Theorem 3} (uniqueness). \emph{Suppose that two functions }$\beta
_{1}\left( x\right) ,\beta _{2}\left( x\right) $\emph{\ satisfying
conditions (\ref{1.1}), (\ref{1.3}). Also, assume that corresponding
functions }$n_{1}\left( x\right) ,n_{2}\left( x\right) $ \emph{satisfy
Assumption of section 2. In addition, assume that these two functions
generate the same function }$f_{1}\left( x,k,y\right) $\emph{\ in (\ref{1.51}%
) and that the function }$f_{1}\left( x,k,y\right) $\emph{\ has the form (%
\ref{6.1}), i.e. the term }$O\left( 1/k\right) $\emph{\ in (\ref{5.5}) is
ignored. Finally, assume that the linearization (\ref{6.7}) is valid. Then }$%
\beta _{1}\left( x\right) \equiv \beta _{2}\left( x\right) $\emph{.}

\section{Reconstruction via the integral geometry}

\label{sec:5}

In this section we present the reconstruction method for the function $\beta
(x)$ from the function $A\left( x,y\right) $ given for all $\left(
x,y\right) \in S\times S.$ This method works for both PISP1 and PISP2.
Indeed, consider two cases:

\textbf{Case 1: PISP1}. Given the function $f_{1}\left( x,k,y\right) $ in (%
\ref{1.51}) and dropping the term $O\left( 1/k\right) $ in (\ref{5.5}), we
obtain the function $A\left( x,y\right) $ via (\ref{6.20}).

\textbf{Case 1: PISP1}. Recall that by (\ref{1.1}) and (\ref{1.3}) $n\left(
x\right) =1$ for $x\in S.$ Hence, in this case (\ref{1.52}), (\ref{1.11})
and (\ref{5.6}) imply that%
\begin{equation}
\lim_{k\rightarrow \infty }\sqrt{f_{2}\left( x,k,y\right) }=A\left(
x,y\right) =\frac{\sqrt{J(x,y)}}{4\pi \tau (x,y)},\text{ }\forall \left(
x,y\right) \in S\times S,  \label{7.1}
\end{equation}%
where the determinant $J(x,y)$ was defined in (\ref{1.91}).

\subsection{Derivation of the problem of the integral geometry}

\label{sec:5.1}

We again use (\ref{6.6}) and the linearization (\ref{6.7}). Rewrite (\ref%
{6.7}) as 
\begin{equation}
\tau (x,y)=|x-y|\Bigg(1+\int\limits_{0}^{1}\beta (y+s(x-y))ds\Bigg).
\label{7.2}
\end{equation}%
Because of (\ref{6.6}), we ignore in our approximate formulas below all
terms containing the square of the integral in (\ref{7.2}) as well as
products of its derivatives. First, using (\ref{1.91}), we find an
approximate formula for the determinant $J(x,y)$ in (\ref{7.1}). Using (\ref%
{1.9a}), we obtain the following approximate formula for $\zeta (x,y)$ 
\begin{equation}
\zeta (x,y)=\frac{(x-y)}{n^{2}(y)}\Bigg(1+2\int\limits_{0}^{1}\beta
(y+s(x-y))ds\Bigg)  \label{7.3}
\end{equation}%
\begin{equation*}
-\frac{|x-y|^{2}}{n^{2}(y)}\int\limits_{0}^{1}\nabla \beta (y+s(x-y))(1-s)ds.
\end{equation*}%
By (\ref{7.3}) we have for $i,j=1,2,3$%
\begin{equation*}
\frac{\partial \zeta _{i}(x,y)}{\partial x_{j}}=\frac{\delta _{ij}}{n^{2}(y)}%
\left[ 1+2\int\limits_{0}^{1}\beta (y+s(x-y))ds\right]
\end{equation*}%
\begin{equation*}
+\frac{2}{n^{2}(y)}\left[ \left( x_{i}-y_{i}\right) \int\limits_{0}^{1}\beta
_{x_{j}}(y+s(x-y))sds-(x_{j}-y_{j})\int\limits_{0}^{1}\beta
_{x_{i}}(y+s(x-y))(1-s)ds\right]
\end{equation*}%
\begin{equation*}
-\frac{|x-y|^{2}}{n^{2}(y)}\int\limits_{0}^{1}\beta
_{x_{i}x_{j}}(y+s(x-y))s(1-s)ds.\quad
\end{equation*}%
Hence, in the $3\times 3$ determinant $J(x,y)=\det \left( \partial \zeta
/\partial x\right) $ the product of diagonal terms dominates the rest of
terms, which should be set to zero because of products of the above
mentioned integrals with the function $\beta $ and its derivatives. Hence,
an approximate formula for $J(x,y)$ is%
\begin{equation*}
J(x,y)=\frac{1}{n^{6}(y)}\left[ 1+6\int\limits_{0}^{1}\beta
(y+s(x-y))ds+2(x-y)\int\limits_{0}^{1}\nabla \beta (y+s(x-y))(2s-1)ds\right.
\end{equation*}%
\begin{equation}
\left. -|x-y|^{2}\int\limits_{0}^{1}\Delta \beta (y+s(x-y))s(1-s)ds\right] .
\label{7.5}
\end{equation}

Note that 
\begin{equation*}
(x-y)\nabla \beta (y+s(x-y))=\frac{\partial \beta (y+s(x-y))}{\partial s}.
\end{equation*}%
Hence, 
\begin{equation*}
(x-y)\int\limits_{0}^{1}\nabla \beta (y+s(x-y))(2s-1)ds=\beta (x)+\beta
(y)-2\int\limits_{0}^{1}\beta (y+s(x-y))ds.
\end{equation*}%
Hence, we obtain 
\begin{equation*}
J(x,y)=1+2\int\limits_{0}^{1}\beta (y+s(x-y))ds
\end{equation*}%
\begin{equation}
-|x-y|^{2}\int\limits_{0}^{1}\Delta \beta (y+s(x-y))s(1-s)ds,\quad (x,y)\in
S\times S.  \label{7.6}
\end{equation}%
Next, by (\ref{7.2})%
\begin{equation}
\frac{1}{\tau \left( x,y\right) }=\frac{1}{|x-y|}\left(
1-\int\limits_{0}^{1}\beta (y+s(x-y))ds\right) ,  \label{7.7}
\end{equation}%
Thus, (\ref{7.1}), (\ref{7.5}) and (\ref{7.6}) imply that 
\begin{equation}
A(x,y)=\frac{1}{4\pi |x-y|}\Bigg(1-\frac{|x-y|^{2}}{2}\int\limits_{0}^{1}%
\Delta \beta (y+s(x-y))s(1-s)ds\Bigg),\>(x,y)\in S\times S.  \label{7.8}
\end{equation}%
Both in (\ref{7.7}) and (\ref{7.8}) we again use \textquotedblleft $=$"
instead of \textquotedblleft $\approx $". Denote $\Delta \beta (x)=q(x)$.
Then we obtain the following problem:

\textbf{The Problem of the Integral Geometry}. \emph{Find the function }$%
q\in C\left( \overline{Y}\right) ,q\left( x\right) =0$\emph{\ for }$x\in
Y\diagdown \Omega ,$\emph{\ assuming that the following integrals }$g(x,y)$%
\emph{\ are given over segments }$L\left( x,y\right) $\emph{\ of straight
lines connecting any two points }$(x,y)\in S\times S,x\neq y$\emph{\ }%
\begin{equation}
|x-y|\int\limits_{0}^{1}q(y+s(x-y))s(1-s)ds=g(x,y),  \label{1.18}
\end{equation}%
\emph{\ }%
\begin{equation*}
g(x,y)=-8\pi A(x,y)+2|x-y|^{-1}.
\end{equation*}

Note that $g(x,y)=g(y,x)$. Since the weight function $s(1-s)$ is present in
the integral (\ref{1.18}), the problem (\ref{1.18}) cannot be solved via the
inversion of the Radon transform as in section 4.2. Therefore, we propose a
different method in subsection 5.2. If the problem (\ref{1.18}) is solved,
then the function $\beta (x)$ can be found via the solution of the Dirichlet
boundary value problem for the Laplace equation, 
\begin{equation*}
\Delta \beta (x)=q(x),\>x\in Y;\quad \beta (x)|_{S}=0.
\end{equation*}

\subsection{Solution of the problem of the integral geometry}

\label{sec:5.2}

We now show that the function $q(x)$ can be reconstructed from (\ref{1.18})
separately in each 2-d cross-section $Q_{z}=\overline{Y}\cap P_{z},z\in
\left( -R,R\right) $ of the ball $Y$. First, we rewrite equation (\ref{1.18}%
) as 
\begin{equation}
\int\limits_{0}^{|x-y|}q\Big(y+s_{1}\frac{(x-y)}{|x-y|}\Big)%
s_{1}(|x-y|-s_{1})ds_{1}=g(x,y),  \label{1.19}
\end{equation}%
\begin{equation*}
\forall (x,y)\in S_{z}\times S_{z},x\neq y,
\end{equation*}%
where $s_{1}$ is the arc length of $L\left( x,y\right) $. Recall that the
plane $P_{z}=\{x_{3}=z\}$. We consider $z\in (-R,R),$ since $q\left(
x\right) =0$ for $\left\vert x\right\vert >R$ (see (\ref{1.3})). In the
plane $P_{z}$ we introduce polar coordinates $r,\varphi $ of the variable
point $\xi =(\xi _{1},\xi _{2})$ as $\xi _{1}=r\cos \varphi ,\xi _{2}=r\sin
\varphi $. We characterize the segment of the straight line $L(x,y)$ passing
through points $x$,$y\in S_{z}$ by the polar coordinates $(\rho ,\alpha )$
of its middle point $(x+y)/2$. Hence, $\left\vert x-y\right\vert =2\sqrt{%
B^{2}-z^{2}-\rho ^{2}}.$ Change variables in the integral (\ref{1.19}) as 
\begin{equation*}
s_{1}\Longleftrightarrow r,s_{1}=\sqrt{B^{2}-z^{2}-\rho ^{2}}-\sqrt{%
r^{2}-\rho ^{2}},
\end{equation*}%
Then $s_{1}(|x-y|-s_{1})=B^{2}-z^{2}-r^{2}.$ The equation of $L(x,y)$ can be
rewritten as 
\begin{equation}
\varphi =\alpha +(-1)^{j}\arccos \frac{\rho }{r},\quad j=1,2,\quad r\geq
\rho ,  \label{100}
\end{equation}%
where $j=1$ corresponds to the part of the straight line $L(x,y)$ between
points $(x+y)/2$ and $x$, and $j=2$ to the rest of $L(x,y)$.

Note that $ds_{1}=-rdr/\sqrt{r^{2}-\rho ^{2}}$. Obviously there exists a
one-to-one correspondence, up to the symmetry mapping $(x,y)\Leftrightarrow
(y,x)$ between pairs $(x,y)\in S_{z}\times S_{z}$ and $(\rho ,\alpha )\in
\left( 0,R\right) \times \left( 0,2\pi \right) $. Denote $q(r\cos \varphi
,r\sin \varphi ,z)=\widetilde{q}(r,\varphi ,z)$ and $g(x,y)=\widetilde{g}%
(\rho ,\alpha ,z)$. Using (\ref{100}), we rewrite equation (\ref{1.19}) as 
\begin{equation}
\sum\limits_{j=1}^{2}\int\limits_{\rho }^{\rho _{0}}\widetilde{q}(r,\alpha
+(-1)^{j}\arccos \frac{\rho }{r},z)\frac{(B^{2}-z^{2}-r^{2})rdr}{\sqrt{%
r^{2}-\rho ^{2}}}=\widetilde{g}(\rho ,\alpha ,z),  \label{1.20}
\end{equation}%
where $\rho _{0}=\sqrt{R^{2}-z^{2}}$. Represent functions $\widetilde{q}%
(r,\varphi ,z)$ and $\widetilde{g}(\rho ,\alpha ,z)$ via Fourier series, 
\begin{equation}
\widetilde{q}(r,\varphi ,z)=\sum\limits_{n=-\infty }^{\infty }\widetilde{q}%
_{n}(r,z)\exp (in\varphi ),  \label{1.22}
\end{equation}%
\begin{equation}
\widetilde{g}(\rho ,\alpha ,z)=\sum\limits_{n=-\infty }^{\infty }\widetilde{g%
}_{n}(\rho ,z)\exp (in\alpha ).  \label{101}
\end{equation}%
Multiplying both sides of (\ref{1.20}) by $\exp (-in\alpha )/(2\pi )$ and
integrating with respect to $\alpha $, we obtain for all $n=0,\pm 1,\pm
2,\ldots $ 
\begin{equation}
\int\limits_{\rho }^{\rho _{0}}\widetilde{q}_{n}(r,z)\cos \Big(n\arccos 
\frac{\rho }{r}\Big)\>\frac{2(B^{2}-z^{2}-r^{2})rdr}{\sqrt{r^{2}-\rho ^{2}}}=%
\widetilde{g}_{n}(\rho ,z).  \label{1.23}
\end{equation}%
Denote 
\begin{equation*}
p_{n}(r,z)=\widetilde{q}_{n}(r,z)(B^{2}-z^{2}-r^{2})r.
\end{equation*}%
Then equation (\ref{1.23}) becomes 
\begin{equation}
\int\limits_{\rho }^{\rho _{0}}p_{n}(r,z)\cos \Big(n\arccos \frac{\rho }{r}%
\Big)\>\frac{2dr}{\sqrt{r^{2}-\rho ^{2}}}=\widetilde{g}_{n}(\rho ,z),\quad
\rho \in (0,\rho _{0}].  \label{1.25}
\end{equation}%
This is the integral equation of the Abel type. To solve equation (\ref{1.25}%
), we apply first the operator $L$ to both sides of (\ref{1.25}), where 
\begin{equation*}
L\left( h\left( \rho \right) \right) \left( s\right) =\frac{1}{\pi }%
\int\limits_{s}^{\rho _{0}}\frac{h(\rho )\rho \,d\rho }{\sqrt{\rho ^{2}-s^{2}%
}},s\in \left( 0,\rho _{0}\right) .
\end{equation*}%
Then changing the limits of the integration, we obtain%
\begin{equation}
\frac{1}{\pi }\dint\limits_{s}^{\rho _{0}}p_{n}(r,z)\left[
\dint\limits_{s}^{r}\frac{2\rho }{\sqrt{\rho ^{2}-s^{2}}\cdot \sqrt{%
r^{2}-\rho ^{2}}}\cos \Big(n\arccos \frac{\rho }{r}\Big)d\rho \right]
dr=L\left( \widetilde{g}_{n}(\rho ,z)\right) \left( s\right) .  \label{102}
\end{equation}%
Change variables in the inner integral (\ref{102}) as%
\begin{equation*}
\rho \Leftrightarrow \theta ,\rho ^{2}=s^{2}\cos ^{2}\left( \theta /2\right)
+r^{2}\sin ^{2}\left( \theta /2\right) .
\end{equation*}%
Then%
\begin{equation*}
2\rho d\rho =\left( r^{2}-s^{2}\right) \sin \theta \cos \theta d\theta ,
\end{equation*}%
\begin{equation*}
\sqrt{\rho ^{2}-s^{2}}\cdot \sqrt{r^{2}-\rho ^{2}}=\left( r^{2}-s^{2}\right)
\sin \theta \cos \theta .
\end{equation*}%
Hence, equation (\ref{102}) can be rewritten as 
\begin{equation}
\int\limits_{s}^{\rho _{0}}p_{n}(r,z)Q_{n}(r,s)dr=L\left( \widetilde{g}%
_{n}(\rho ,z)\right) \left( s\right) ,  \label{1.26}
\end{equation}%
\begin{equation*}
Q_{n}(r,s)=\frac{1}{\pi }\int\limits_{0}^{\pi }\cos \left( n\arccos \frac{%
\sqrt{r^{2}\cos ^{2}(\theta /2)+s^{2}\sin ^{2}(\theta /2)}}{r}\right)
d\theta .
\end{equation*}%
We have $Q_{n}(s,s)=1$. Hence, differentiating (\ref{1.26}) with respect to $%
s$, we obtain Volterra integral equation of the second kind 
\begin{equation}
p_{n}(s,z)-\int\limits_{s}^{\rho _{0}}p_{n}(r,z)T_{n}(r,s)dr=-\frac{\partial 
}{\partial s}\left[ L\left( \widetilde{g}_{n}(\rho ,z)\right) \left(
s\right) \right] ,s\in \left( 0,\rho _{0}\right) ,  \label{1.28}
\end{equation}%
\begin{eqnarray}
T_{n}(r,s) &=&\frac{ns}{\pi \sqrt{r^{2}-s^{2}}}\int\limits_{0}^{\pi }\sin
\left( n\arccos \frac{\sqrt{r^{2}\cos ^{2}(\theta /2)+s^{2}\sin ^{2}(\theta
/2)}}{r}\right)  \label{1.29} \\
&&\times \frac{\sin (\theta /2)d\theta }{\sqrt{r^{2}\cos ^{2}(\theta
/2)+s^{2}\sin ^{2}(\theta /2)}}.  \notag
\end{eqnarray}%
It follows from (\ref{1.29}) that the kernel of the Volterra integral
equation (\ref{1.28}) has the form%
\begin{equation*}
T_{n}(r,s)=\frac{\widetilde{T}_{n}(r,s)}{\sqrt{r^{2}-s^{2}}},
\end{equation*}
where the function $\widetilde{T}_{n}(r,s)$ is continuous for $0\leq s\leq
r\leq \rho _{0}$. Therefore, it follows from the theory of Volterra integral
equations of the second kind that for each $z\in \left( -R,R\right) $ there
exists a solution $p_{n}(s,z)\in C\left[ 0,\rho _{0}\right] $ of equation (%
\ref{1.28}) and this solution is unique. Furthermore, it is well known from
that theory that equation (\ref{1.28}) can be solved iteratively as%
\begin{equation}
p_{n}^{0}(s,z)=-\frac{\partial }{\partial s}\left[ L\left( \widetilde{g}%
_{n}(\rho ,z)\right) \left( s\right) \right] ,  \label{1.30}
\end{equation}%
\begin{equation}
p_{n}^{k}(s,z)=\int\limits_{s}^{\rho _{0}}p_{n}^{k-1}(r,z)T_{n}(r,s)dr-\frac{%
\partial }{\partial s}\left[ L\left( \widetilde{g}_{n}(\rho ,z)\right)
\left( s\right) \right] ,k=1,2,...  \label{1.32}
\end{equation}%
and this process converges in the space $C\left[ 0,\rho _{0}\right] $ to the
solution $p_{n}(s,z)$ of equation (\ref{1.28}). Formulae (\ref{1.30}) and (%
\ref{1.32}) finish our second reconstruction procedure.

As a corollary, we formulate the following uniqueness theorem, which follows
immediately from the above reconstruction process.

\textbf{Theorem 4 }(uniqueness)\textbf{.} \emph{Assume that the function }$%
A\left( x,y\right) $\emph{\ has its approximate form (\ref{7.8}). Suppose
that this function is given for all }$(x,y)\in S\times S.$\emph{\ Then there
exists at most one function }$\beta \in C^{2}\left( \overline{Y}\right)
,\beta \left( x\right) =0$\emph{\ in }$Y\diagdown \Omega $\emph{\ which is
involved in (\ref{7.8}). In particular, PISP2 has at most one solution as
long as the function }$A\left( x,y\right) $\emph{\ is as in (\ref{7.8}). The
same is true for PISP1, if, in addition, the term }$O\left( 1/k\right) $%
\emph{\ is dropped in (\ref{5.5}).}\hfill $\Box $

\end{document}